\documentclass[12pt,prd,tightenlines,nofootinbib]{revtex4}
\usepackage{bm}
\usepackage{graphics}
\usepackage{rotating}
\usepackage{epsfig}
\begin{document}

\title{Relativistic description of heavy tetraquarks}
\author{D. Ebert$^1$, R. N. Faustov$^{2}$  and V. O. Galkin$^{2}$}
\affiliation{$^1$ Institut f\"ur Physik, Humboldt--Universit\"at zu Berlin,
Newtonstr. 15, D-12489  Berlin, Germany\\
$^2$ Dorodnicyn Computing Centre, Russian Academy of Sciences,
  Vavilov Str. 40, 119991 Moscow, Russia}

\begin{abstract}
 The masses of the ground state and excited heavy tetraquarks with hidden
charm and bottom are calculated within  the relativistic
diquark-antidiquark picture. The dynamics of the light quark in a
heavy-light diquark is treated completely relativistically. The
diquark structure is taken into account by calculating
the diquark-gluon form factor. New experimental data on charmonium-like states
above the open charm threshold are discussed. The obtained results
indicate that  $X(3872)$, $Y(4260)$, $Y(4360)$, $Z(4433)$ and
$Y(4660)$ can be tetraquark states with hidden charm.  
\end{abstract}

\pacs{12.40.Yx, 14.40.Gx, 12.39.Ki}

\maketitle

Recently the significant experimental progress has been achieved in heavy
hadron spectroscopy. Several  new charmonium-like
states, such as $X(3872)$, $Y(4260)$, $Y(4360)$, $Y(4660)$, $Z(4430)$,
etc., were 
observed \cite{pakhlova} which cannot be simply accommodated in the 
quark-antiquark ($q\bar q$) picture. These  states
can be considered as indications of the possible existence of
exotic multiquark states \cite{jm,bk}.
Here we briefly review our recent results for the masses of
heavy tetraquarks in  the framework of the relativistic quark
model based on the quasipotential approach in quantum chromodynamics.
We use the  diquark-antidiquark
approximation to reduce a complicated relativistic 
four-body problem to the subsequent more simple two-body
problems. The first step consists in the calculation of the masses, wave
functions and form factors of the diquarks, composed from light and heavy
quarks. At the final step, a heavy tetraquark is considered to be a
bound diquark-antidiquark system. It is 
important to emphasize that we do not consider a diquark as a point
particle but explicitly take into account its structure by calculating
the form factor of the diquark-gluon interaction in terms of the
diquark wave functions.

In the quasipotential approach the two-particle bound state with the mass
$M$ and masses of the constituents $m_{1,2}$ in momentum
representation is described by the wave
function $\Psi({\bf p})$ 
satisfying the quasipotential equation of the Schr\"odinger type
\begin{equation}
\label{quas}
{\left(\frac{b^2(M)}{2\mu_{R}}-\frac{{\bf
p}^2}{2\mu_{R}}\right)\Psi_{d,T}({\bf p})} =\int\frac{d^3 q}{(2\pi)^3}
 V_{d,T}({\bf p,q};M)\Psi_{d,T}({\bf q}),
\end{equation}
where the relativistic reduced mass is
\[
\mu_{R}=\frac{M^4-(m^2_1-m^2_2)^2}{4M^3},\]
and the on-mass-shell relative momentum squared
\[ {b^2(M) }
=\frac{[M^2-(m_1+m_2)^2][M^2-(m_1-m_2)^2]}{4M^2}.
\]
The subscript $d$ refers to the diquark and $T$ refers to the
tetraquark composed of a diquark and antidiquark.
The explicit expressions for the corresponding quasipotentials $V_{d,T}({\bf p,q};M)$  can be
found in Ref.~\cite{tetr}.

At the first step, we calculate the masses and form factors of the
light and heavy
diquarks. As it is well known, the light quarks are highly
relativistic, which makes the $v/c$ expansion inapplicable and thus,
a completely relativistic treatment of the light quark dynamics is required. To achieve this goal we closely follow our consideration
of the mass spectra of light 
mesons and adopt the same procedure to make the relativistic
potential local by replacing
$\epsilon_{1,2}(p)=\sqrt{m_{1,2}^2+{\bf p}^2}\to E_{1,2}=(M^2-m_{2,1}^2+m_{1,2}^2)/2M$. 
Solving numerically the quasipotential equation (\ref{quas}) with the
complete relativistic potential,  which depends on the
diquark mass in a complicated highly nonlinear way \cite{hbar}, we get
the diquark masses and wave functions. In order to determine the
diquark interaction with the gluon field, which 
takes into account the diquark structure, we
calculate the corresponding matrix element of the quark
current between diquark states. Such calculation leads to the
emergence of the form factor $F(r)$ entering the vertex of the
diquark-gluon interaction \cite{hbar}. This form factor is expressed
through the overlap integral of the diquark wave functions. Our
estimates show that it can be approximated  with a 
high accuracy by the expression 
\begin{equation}
  \label{eq:fr}
  F(r)=1-e^{-\xi r -\zeta r^2}.
\end{equation}
The values of the masses and parameters $\xi$ and $\zeta$ for heavy-light
scalar diquark $[Q,q]$ and axial vector diquark $\{Q,q\}$ ground states are
given in Table~\ref{tab:dmass}.

\begin{table}
  \caption{Masses $M$ and form factor  parameters of heavy-light
    diquarks. $S$ and $A$ 
    denote scalar and axial vector diquarks antisymmetric $[Q,q]$ and
    symmetric $\{Q,q\}$ in flavour, respectively. }
  \label{tab:dmass}
\begin{ruledtabular}
\begin{tabular}{cccccccc}
Quark& Diquark&  
\multicolumn{3}{l}{\underline{\hspace{2.5cm}$Q=c$\hspace{2.5cm}}}
\hspace{-3.4cm}
&\multicolumn{3}{l}{\underline{\hspace{2.5cm}$Q=b$\hspace{2.5cm}}}
\hspace{-3.4cm} \\
content &type & $M$ (MeV)&$\xi$ (GeV)&$\zeta$ (GeV$^2$)  & $M$
(MeV)&$\xi$ (GeV)&$\zeta$ (GeV$^2$) \\
\hline
$[Q,q]$& $S$ & 1973& 2.55 &0.63 & 5359 &6.10 & 0.55 \\
$\{Q,q\}$& $A$ & 2036& 2.51 &0.45 & 5381& 6.05 &0.35 \\
$[Q,s]$ & $S$& 2091& 2.15 & 1.05 & 5462 & 5.70 &0.35 \\
$\{Q,s\}$& $A$ & 2158&2.12& 0.99 & 5482 & 5.65 &0.27
  \end{tabular}
\end{ruledtabular}
\end{table}

At the final step, we calculate the masses of heavy tetraquarks 
considered as the bound states of a heavy-light diquark and
antidiquark. In this picture of heavy tetraquarks
both scalar $S$ (asymmetric in flavour
$[Qq]_{S=0}=[Qq]$) and axial vector $A$ (symmetric in flavour
$[Qq]_{S=1}=\{Qq\}$) diquarks are considered. Therefore we get the
following structure of the $[Qq][\bar Q\bar q']$  ground ($1S$) states
($C$ is defined only for $q=q'$): 
\begin{itemize}
\item Two states with $J^{PC}=0^{++}$:
\begin{eqnarray*}
&&X(0^{++})=[Qq]_{S=0}[\bar Q\bar q']_{S=0}\\
&&X(0^{++}{}')=[Qq]_{S=1}[\bar Q\bar q']_{S=1}
\end{eqnarray*}
\item Three states with $J=1$:
\begin{eqnarray*}
&&X(1^{++})=\frac1{\sqrt{2}}([Qq]_{S=1}[\bar Q\bar q']_{S=0}+[Qq]_{S=0}[\bar Q\bar
  q']_{S=1})\\
&&X(1^{+-})=\frac1{\sqrt{2}}([Qq]_{S=0}[\bar Q\bar q']_{S=1}-[Qq]_{S=1}[\bar Q\bar
  q']_{S=0})\\
&&X(1^{+-}{}')=[Qq]_{S=1}[\bar Q\bar q']_{S=1}
\end{eqnarray*}
\item  One state with $J^{PC}=2^{++}$:
$$X(2^{++})=[Qq]_{S=1}[\bar Q\bar q']_{S=1}.$$
\end{itemize}
The orbitally excited ($1P,1D\dots$) states are
constructed analogously. As we see a very rich spectrum of tetraquarks
emerges. However the number of states in the considered
diquark-antidiquark picture is significantly less than in the genuine
four-quark approach.

The diquark-antidiquark model of heavy tetraquarks predicts
\cite{mppr} the existence of the flavour 
$SU(3)$ nonet of states with hidden
charm or beauty ($Q=c,b$): four tetraquarks
($[Qq][\bar Q\bar q]$, $q=u,d$) with neither open 
or hidden strangeness, which have
electric charges 0 or $\pm 1$ and isospin 0 or 1; 
four tetraquarks ($[Qs][\bar Q\bar q]$
and  $[Qq][\bar Q\bar s]$, $q=u,d$) with open strangeness ($S=\pm 1$),
which have electric charges 0 or $\pm 1$ and isospin $\frac12$; 
one tetraquark
($[Qs][\bar Q\bar s]$) with hidden strangeness and zero electric
charge. 
Since in our model we neglect the mass difference of $u$ and
$d$ quarks and electromagnetic interactions, corresponding tetraquarks
will be degenerate in mass. A more 
detailed analysis \cite{mppr}
predicts that such mass differences can be of a few MeV so
that the
isospin invariance is broken for the $[Qq][\bar Q\bar q]$ mass
eigenstates and thus in their strong decays.  
The (non)observation of such states will be a crucial test of the
tetraquark model. 

\begin{table}
  \caption{Masses of charm diquark-antidiquark ground ($1S$) states  (in MeV). $S$ and $A$
    denote scalar and axial vector diquarks. }
  \label{tab:cmass}
\begin{ruledtabular}
\begin{tabular}{ccccc}
State& Diquark &
\multicolumn{3}{l}{\underline{\hspace{3.6cm}Mass\hspace{3.6cm}}} 
\hspace{-5.5cm} \\
$J^{PC}$ & content& $cq\bar c\bar q$ &$cs\bar c\bar s$ & $cs\bar c\bar
q/ cq\bar c\bar s$ \\
\hline
$0^{++}$ & $S\bar S$ & 3812 & 4051 & 3922\\
$1^{+\pm}$ & $(S\bar A\pm \bar S A)/\sqrt2$& 3871& 4113 & 3982\\
$0^{++}$& $A\bar A$ & 3852 & 4110& 3967\\
$1^{+-}$& $A\bar A$ & 3890 & 4143& 4004\\
$2^{++}$& $A\bar A$ & 3968 & 4209&4080\\
 \end{tabular}
\end{ruledtabular}
\end{table}

\begin{table}
  \caption{Thresholds for open charm decays and nearby hidden-charm
    thresholds.} 
  \label{tab:cthr}
\begin{ruledtabular}
\begin{tabular}{cccccc}
Channel& Threshold (MeV)&Channel& Threshold (MeV)&Channel& Threshold
(MeV)\\ 
\hline
$D^0\bar D^0$& 3729.4 &$D_s^+ D_s^-$& 3936.2&$D^0 D_s^\pm$& 3832.9\\
$D^+D^-$& 3738.8& $\eta' J/\psi$& 4054.7& $D^\pm D_s^\mp$ & 3837.7\\
$D^0\bar D^{*0}$ & 3871.3& $D_s^\pm D_s^{*\mp}$& 4080.0&$D^{*0} D_s^\pm$ &
3975.0\\
$\rho J/\psi$& 3872.7& $\phi J/\psi$ & 4116.4&$D^{0}D^{*\pm}_s$ &
3976.7\\
$D^\pm D^{*\mp}$ &3879.5 &$D^{*+}_sD^{*-}_s$& 4223.8& $K^{*\pm}J/\psi$ &
3988.6\\
$\omega J/\psi$& 3879.6 & & &  $K^{*0}J/\psi$ & 3993.0\\
$D^{*0}\bar D^{*0}$ & 4013.6 & & &$D^{*0} D_s^{*\pm}$ & 4118.8   
\end{tabular}
\end{ruledtabular}
\end{table}

The calculated  masses of the heavy tetraquark ground ($1S$) states
and the corresponding open charm  and bottom thresholds are given in
Tables~~\ref{tab:cmass}-\ref{tab:bthr}. We find that all $S$-wave
tetraquarks with hidden bottom lie considerably below
open bottom thresholds and thus they should be narrow states which can
be observed experimentally. This prediction significantly differs from the
molecular picture  where bound $B-\bar B^*$ states are
expected to lie very close (only few MeV below) to the corresponding
thresholds. 

\begin{table}
  \caption{Masses of bottom diquark-antidiquark ground ($1S$) states (in MeV). $S$ and $A$
    denote scalar and axial vector diquarks. }
  \label{tab:bmass}
\begin{ruledtabular}
\begin{tabular}{ccccc}
State& Diquark &
\multicolumn{3}{l}{\underline{\hspace{3.7cm}Mass\hspace{3.7cm}}} 
\hspace{-5.5cm} \\
$J^{PC}$ & content& $bq\bar b\bar q$ &$bs\bar b\bar s$ & $bs\bar b\bar
q/ bq\bar b\bar s$ \\
\hline
$0^{++}$ & $S\bar S$ & 10471 & 10662 & 10572\\
$1^{+\pm}$ & $(S\bar A\pm \bar S A)/\sqrt2$& 10492& 10682 & 10593\\
$0^{++}$& $A\bar A$ & 10473 & 10671& 10584\\
$1^{+-}$& $A\bar A$ & 10494 & 10686& 10599\\
$2^{++}$& $A\bar A$ & 10534 & 10716& 10628\\
 \end{tabular}
\end{ruledtabular}
\end{table}

\begin{table}
  \caption{Thresholds for open bottom decays.}
  \label{tab:bthr}
\begin{ruledtabular}
\begin{tabular}{cccccc}
Channel& Threshold (MeV)&Channel& Threshold (MeV)&Channel& Threshold
(MeV)\\ 
\hline
$B\bar B$ & 10558& $B_s^+B_s^-$ &10739 & $B B_s$ &10649\\
$B\bar B^*$ & 10604 &$B_s^\pm B_s^{*\mp}$ & 10786 &$B^* B_s$ & 10695\\
$B^*\bar B^*$ & 10650 &$B_s^{*+} B_s^{*-}$ & 10833 &$B^* B_s^*$& 10742  
\end{tabular}
\end{ruledtabular}
\end{table}

The situation  in the hidden charm sector is considerably more
complicated, since most of the tetraquark states are predicted to lie
either above or only slightly below corresponding open charm
thresholds. This difference is the consequence of the fact that the
charm quark mass is substantially smaller than the bottom quark
mass. As a result the binding energies in the
charm sector are
significantly smaller than those in the bottom sector.

\begin{table}
  \caption{Comparison of theoretical predictions for  the masses of
    the ground and excited
   charm diquark-antidiquark states $cq\bar c\bar q$ (in MeV) and
   possible experimental candidates.} 
  \label{tab:cemass}
\begin{ruledtabular}
\begin{tabular}{ccccccc}
State&Diquark&
\multicolumn{3}{l}{\underline{\hspace{2.8cm}Theory\hspace{2.8cm}}}
& \multicolumn{2}{l}{\underline{\hspace{1.7cm}Experiment
    \hspace{1.7cm}}} 
\hspace{-1.5cm}  \\
$J^{PC}$&content &EFG & Maiani et al. &Maiani et al. ($cs\bar c\bar s$) 
&state& mass\\
\hline
$1S$\\
$0^{++}$&$S\bar S$ & 3812 & 3723& & &\\
$1^{++}$&$(S\bar A+ \bar S A)/\sqrt2$ & 3871& 3872$^\dag$&
&$\left\{\begin{array}{l} X{(3872)}\\ X{(3876)}\end{array}\right.$ 
&$\left\{\begin{array}{l}{3871.4}\pm{0.6}\\
{3875.4}\pm{0.7}^{+1.2}_{-2.0}\end{array}\right.$ \\
$1^{+-}$&$(S\bar A- \bar S A)/\sqrt2$ & 3871& 3754&& &\\
$0^{++}$&$A\bar A$& 3852 & 3832&& &\\
$1^{+-}$&$A\bar A$& 3890 & 3882&& &\\
$2^{++}$&$A\bar A$& 3968 & 3952&&$Y$(3943)&$\left\{\begin{array}{l}3943\pm11\pm13 \\3914.3^{+4.1}_{-3.8}\end{array}\right.$\\
$1P$\\
$1^{--}$&$S\bar S$&4244 & &4330$\pm$70&$Y$(4260)
&$\left\{\begin{array}{l}{4259}\pm{8}^{+2}_{-6}\\
   {4247}\pm{12}^{+17}_{-32}\end{array}\right.$\\
$1^{--}$&$(S\bar A- \bar S A)/\sqrt2$ &4284& &&$Y$(4260) &
4283$^{+17}_{-16}\pm$4 \\
$1^{--}$&$A\bar A$& 4277 \\
$1^{--}$&$A\bar A$& 4350& &  &$Y$(4360) & 4361$\pm$9$\pm$9\\
$2S$\\
$1^{+\pm}$&$(S\bar A\pm \bar S A)/\sqrt2$ &4431& &
&$Z$(4430)&4433$\pm$4$\pm$1\\
$0^{++}$&$A\bar A$&4434& \\
$1^{+-}$&$A\bar A$&4461& $\sim$ 4470\\
$2P$\\
$1^{--}$&$S\bar S$&4666& & & $Y$(4660)& 4664$\pm$11$\pm$5\\ 
\end{tabular}
 \end{ruledtabular}
\flushleft{${}^\dag$ input}
\end{table}

In Table~\ref{tab:cemass} we compare our results (EFG \cite{tetr}) for
the masses of the ground and excited charm 
diquark-antidiquark bound states with the predictions of
Ref.~\cite{mppr,mpr,mpprY,mprZ} and with the masses of the recently observed
excited charmonium-like states \cite{pakhlova}. 
We  assume that the excitations  occur only inside the
diquark-antidiquark bound system. Possible excitations of diquarks are
not considered. Our calculation of the heavy baryon masses supports
such scheme \cite{hbar}.
In this table we give our predictions only for some of the masses of the
orbitally and radially excited states for which possible experimental
candidates are available. 
The differences in some of the presented theoretical mass values can
be attributed to the substantial distinctions in the used
approaches. We describe the diquarks dynamically as 
quark-quark bound 
systems and  calculate their masses and form factors, while in
Ref.\cite{mppr}  they
are treated only phenomenologically. Then we consider the tetraquark
as purely the 
diquark-antidiquark bound system.  In distinction Maini et al. 
consider a
hyperfine interaction between all quarks  which, e.g., causes the
splitting of $1^{++}$ and $1^{+-}$ states arising from the $SA$
diquark-antidiquark compositions.  
From Table~\ref{tab:cemass} we see that our dynamical 
calculation supports  the assumption \cite{mppr} that $X(3872)$ can be
the axial vector 
$1^{++}$ tetraquark state composed from the scalar and axial vector
diquark and antidiquark in the relative $1S$ state. Recent Belle and
BaBar results indicate the existence of a second $X(3875)$ particle a
few MeV above $X(3872)$. This state could be naturally identified
with the second neutral particle predicted by the tetraquark model \cite{mpr}.  
On the other hand, in our model 
the lightest scalar $0^{++}$ tetraquark is
predicted to be above the open charm threshold $D\bar D$
and thus to be broad, while in the model \cite{mppr} it lies 
few MeV below this threshold, and thus is predicted to be narrow. Our
$2^{++}$ state also lies higher than the one in Ref.\cite{mppr},
thus making the interpretation of this state as $Y(3943)$ less
probable especially if one averages the original Belle mass with the recent
BaBar value wich is somewhat lower.  

The recent discovery of the $Y(4260)$, $Y(4360)$ and $Y(4660)$
indicates an excess of the expected charmonium $1^{--}$
states \cite{pakhlova}. The absence of open charm production is also inconsistent with
a conventional $c\bar c$ explanation.  Maini et al. \cite{mpprY} argue that
$Y(4260)$ is the  $1^{--}$ $1P$ state of the charm-strange
diquark-antidiquark tetraquark.
We find that $Y(4260)$ cannot be interpreted in this way, since the mass
of such $([cs]_{S=0}[\bar c\bar s]_{S=0})$ tetraquark  is found to
be $\sim 200$ MeV higher. A more natural
tetraquark interpretation could be the $1^{--}$ $1P$ state  $([cq]_{S=0}[\bar
c\bar q]_{S=0})$ ($S\bar S$) which mass is predicted in our model to be 
close to the mass of  $Y(4260)$ (see Table~\ref{tab:cemass}). 
Then the $Y(4260)$ would decay dominantly into $D\bar D$ pairs.
The other possible interpretations of  $Y(4260)$ are the $1^{--}$ $1P$ states of
$(S\bar A- \bar S A)/\sqrt2$ and $A\bar A$ tetraquarks which predicted masses
have close values. These additional tetraquark states could be
responsible for the mass difference of $Y(4260)$  observed in different
decay channels. As we see from Table~\ref{tab:cemass} the recently
discovered resonances  $Y(4360)$ and 
$Y(4660)$ in the $e^+e^-\to\pi^+\pi^-\psi'$ cross section can be
interpreted as the excited $1^{--}$ $1P$ ($A\bar A$) and $2P$ ($S\bar S$)
tetraquark states, respectively. 

Very recently the Belle Collaboration reported observation of a
relatively narrow enhancement in the $\pi^+\psi'$ invariant mass
distribution in the $B\to K \pi^+\psi'$ decay \cite{pakhlova}. This new resonance,
$Z^+(4430)$, is unique among other exotic meson candidates since it
has a non-zero electric charge. Different theoretical interpretations
were suggested \cite{pakhlova}. Maiani et al. \cite{mprZ} give  qualitative arguments that
the $Z^+(4430)$ could be the first radial excitation ($2S$) of a diquark-antidiquark
$X^+_{u\bar d}(1^{+-}; 1S)$ state ($A\bar A$) with mass 3882 MeV. Our calculations
indicate that the $Z^+(4430)$ can indeed be the $1^{+-}$ $2S$
$[cu][\bar c \bar d]$ tetraquark state. It is the first radial
excitation of the  ground state $(S\bar A- \bar S A)/\sqrt2$, which has the
same mass as $X(3872)$. 

In summary, we calculated the masses of heavy tetraquarks with hidden
charm and bottom in the diquark-antidiquark picture. In contrast to
previous phenomenological treatments we used the dynamical approach
based on the relativistic quark model. Both diquark and tetraquark
masses were obtained by numerical solution of the quasipotential
equation with the corresponding relativistic 
potentials. The diquark structure was also taken into account with the help of
the diquark-gluon form factor expressed in terms of diquark wave
functions. It is important to emphasize  
that, in our analysis, we did not introduce any free adjustable
parameters but used their fixed values from our previous considerations
of heavy and light meson properties. It was found that the $X(3872)$,
$Y(4260)$, $Y(4360)$, $Z(4433)$ and $Y(4660)$ exotic meson candidates
can be tetraquark states with hidden charm. The ground states of
bottom tetraquarks are predicted to have 
masses below the 
open bottom threshold and thus should be narrow.

The authors are grateful to A. Badalian, A. Kaidalov, V. Matveev,
G. Pakhlova, P. Pakhlov, A. Polosa and 
V. Savrin  for support and discussions.  This work was supported in part by the {\it Deutsche
Forschungsgemeinschaft} under contract Eb 139/2-4 and by the {\it Russian
Foundation for Basic Research} under Grant No.05-02-16243.




\end{document}